\documentclass[11pt,A4paper,twoside]{article}
\newif\ifdraft \global\drafttrue
\def\production{\global\draftfalse}
\production
\usepackage[T1]{fontenc}
\usepackage[latin1]{inputenc}
\usepackage[american]{babel}
\usepackage{a4wide}
\usepackage{times}
\usepackage{graphicx}
\usepackage{theorem}
\usepackage[colorlinks=true,hyperindex=true]{hyperref}
\usepackage{color}
\usepackage{fancyhdr}
\usepackage{latexsym}
\usepackage{amsmath}
\usepackage{bbm}
\usepackage{amssymb}
\usepackage{amsfonts}
\usepackage{enumerate}
\usepackage{cancel}
\usepackage{stmaryrd}
\ifdraft
\usepackage{showkeys}
\fi
\newcounter{smallarabics}

\newcounter{smallroman}

\newcommand{\ben}{\begin{enumerate}[{\rm (1)}]}
\newcommand{\een}{\end{enumerate}}

\newtheorem{theoreme}{Theorem }[section]
\newtheorem{proposition}[theoreme]{Proposition}
\newtheorem{lemma}[theoreme]{Lemma}
\newtheorem{definition}[theoreme]{Definition}
\newtheorem{corollary}[theoreme]{Corollary}

\def\bep{\begin{proposition}}
\def\eep{\end{proposition}}

\def\bel{\begin{lemma}}
\def\eel{\end{lemma}}
\def\bet{\begin{theoreme}}
\def\eet{\end{theoreme}}
\def\bed{\begin{definition}}
\def\eed{\end{definition}}
\def\bec{\begin{corollary}}
\def\eec{\end{corollary}}

\def\rr{{\mathbb R}}

\def\cc{{\mathbb C}}
\def\nn{{\mathbb N}}

\def\textsl{{}}

\def\Re{{\rm Re}\,}

\newcommand{\slim}{\mathop{\mathrm{s-lim}}\limits}

\def\c0inf{C_0^\infty}

\def\proof{\noindent {\bf Proof.}\ \ }

\def\cR{{\cal R}}

\def\PP{\mathbb{P}}

\def\QQ{\mathbb{Q}}
\def\kB{{k_\mathrm{B}}}
\def\Int{{\rm int}}
\def\Ext{{\rm ext}}

\def\i{{\rm i}}

\newcommand{\beq}{\begin{equation}}
\newcommand{\eeq}{\end{equation}}
\newcommand{\bear}[1]{\begin{array}{#1}}
\newcommand{\ear}{\end{array}}

\newcommand{\e}{\mathrm{e}}
\renewcommand{\i}{\mathrm{i}}

\renewcommand{\d}{\mathrm{d}}

\def\qed{$\Box$\medskip}

\def\cC{{\cal C}}
\def\cT{{\cal T}}

\def\cL{{\cal L}}
\def\cN{{\cal N}}

\def\cK{{\cal K}}

\def\12{\frac{1}{2}}

\def\e{{\rm e}}

\def\d{{\rm d}}
\def\Ran{{\rm Ran}\,}

\def\cR{{\cal R}}

\def\P{\mathbb P}

\def\cO{{\cal O}}

\def\vg{\mathrm{vg}}

\begin{document}
\def\today{}
\title{Energy statistics in open harmonic networks}
\author{ Tristan  Benoist$^{1}$, Vojkan Jak\v{s}i\'c$^{2}$,  Claude-Alain  Pillet$^{3}$
\\ \\ \\
$^1$Institut de Mathématiques de Toulouse, Équipe de Statistique et Probabilités,\\
 Université Paul
Sabatier, 31062 Toulouse Cedex 9, France
\\ \\
$^2$Department of Mathematics and Statistics, 
McGill University, \\
805 Sherbrooke Street West, 
Montreal,  QC,  H3A 2K6, Canada
\\ \\
$^3$Aix Marseille Univ, Université de Toulon, CNRS, CPT, Marseille, France  
}

\maketitle

{\small
{\bf Abstract.} We relate the large time asymptotics of the energy statistics in
open harmonic networks to the variance-gamma distribution and prove a full Large
Deviation Principle. We consider both Hamiltonian and stochastic dynamics, the
later case including  electronic RC networks. We compare our theoretical
predictions with the experimental data obtained by Ciliberto et al.~\cite{CINT}.}

\thispagestyle{empty}


\section{Introduction} 
\label{sec-intro}
This note grew out of a set of remarks  concerning energy fluctuations in
thermal equilibrium for open harmonic Hamiltonian or stochastic networks made 
in~\cite{JPS1,JPS2} (see Remark 5 in Section 3.2 of~\cite{JPS1} and Remark 3.25 in
Section 3.6 of~\cite{JPS2}) and our attempts to interpret the experimental
results of~\cite{CINT} in this  context. From a theoretical point of view, it
turned out  that for this very specific question one can go far beyond the large
deviation results of~\cite{JPS1,JPS2}  and the above mentioned remarks: one can
compute the statistics of energy fluctuations in steady state network in a
closed form in terms of the variance-gamma distribution. Equally surprisingly,
for some specific electronic RC-circuit this statistics and the resulting
variance-gamma distribution were precisely measured in experiments of Ciliberto,
Imparato, Naert, and Tanase~\cite{CINT}, although the  result was not explicitly
reported (the focus of the experiment was the measurement of the non-equilibrium
entropy production statistics). The purpose of this note is to comment on the
theoretical derivation of the heat and work statistics in harmonic networks and
RC-circuits, and  in the latter case  compare the theoretical prediction with
experimental results of~\cite{CINT}.

To describe more precisely the line of thought that has led to this note,
consider a stationary stochastic process $Z_t$, $t\geq0$, and set 
\[
Q_t=Z_t- Z_0.
\]
Obviously, $\langle Q_t\rangle=0$, where $\langle\,\cdot\,\rangle$ denotes 
expectation, and one may ask about order one fluctuations of $\frac{1}{t}Q_t$.
A standard way to approach the respective large deviations problem is to study 
the limit
\[
q(\alpha)=\lim_{t\rightarrow \infty}\frac{1}{t}
\log\left\langle\e^{\alpha Q_t}\right\rangle
\]
for $\alpha$ in $\rr$. Of course, if $Z_0$ is almost surely bounded, then 
$\frac{1}{t}Q_t$ has no order one fluctuations and $q(\alpha)=0$ for all 
$\alpha$. However, in some physically important cases with unbounded $Z_0$, one 
has  $q(\alpha)=0$ for $|\alpha|\leq \alpha_c$ and $q(\alpha)=\infty$ for 
$|\alpha|>\alpha_c$, where the critical $\alpha_c$ is finite. This was precisely 
the case in the above mentioned remarks in~\cite{JPS1,JPS2}, where 
$Z_t- Z_0$ was related to the energy flux  across  a finite part of harmonic 
Hamiltonian network in thermal equilibrium. If the  Large Deviations Principle 
for $\frac{1}{t}Q_t$ is controlled by the Legendre transform 
$I(\theta)=\alpha_c|\theta|$  of $q(\alpha)$, then for any open set 
$O\subset \rr$, 
\beq
\lim_{t\rightarrow\infty}\frac{1}{t}\log\QQ_t(tO)
=-\inf_{\theta\in{O}}I(\theta),
\label{LDP-i}
\eeq
where $\QQ_t$ denotes the probability distribution of $Q_t$. The rigorous
justification of this step in~\cite{JPS1, JPS2} required an application of a
somewhat subtle extension of the G\"artner-Ellis theorem. The conclusion of the
remarks is that a physically important quantity like energy variation per unit
time  in thermal equilibrium, which has expectation zero, may exhibit order one
fluctuations due to the fact that the phase space space is unbounded.

The experimental results  of~\cite{CINT} indicated that a similar phenomenon
happens and can be precisely measured in a suitable electronic RC-circuit. In
examining the connection we realized that the large deviation
result~\eqref{LDP-i} can be considerably refined and that one can express the
large time statistics of the energy fluctuations in a finite part of an open
harmonic network in closed form in terms of the variance-gamma distribution. That
in turn reflected on the experimental findings of~\cite{CINT}. In their setting,
the large time statistics of energy change turns into the large time statistics
of the total power dissipated in the resistors and is also described by the
variance-gamma distribution. This allowed us to compare the theoretical
prediction of the variance-gamma distribution with the experimental results 
of~\cite{CINT}. The comparison is presented in Subsection~\ref{sec-CINT}. The
theoretical prediction is nearly perfectly matched with experimental results.

Although the  mathematical analysis presented in this note is elementary, the
conclusions shed a new light on some important aspects of statistical mechanics
of open system and can be viewed as a fine statistical characterization of the
quantities entering the first law of thermodynamic for finite subsystems of
extended harmonic systems. For example, we show that the total kinetic energy of
a finite subsystem verifies a universal LDP~\eqref{LDP-i} with
$\alpha_c=(k_BT)^{-1}$. These points requires further study and we plan to
return to them in the future.

The note is organized as follows. In Section~\ref{sec-prob} we review the
derivation of the variance-gamma distribution and describe its basic properties.
Hamiltonian harmonic networks are discussed in Section~\ref{sec-det}.
Stochastic harmonic networks and electronic RC-circuits are  discussed
in Section~\ref{sec-stoch}.

\bigskip\noindent
{\bf Acknowledgments.} We are grateful to Sergio Ciliberto and David Stephens
for useful discussions. We also wish to thank Sergio Ciliberto for making
available to us the raw data of the experiment reported in~\cite{CINT} which
allowed one of us (T.B.) to analyze the data and  plot the
Figure~\ref{fig:CINT_graph} in Section~\ref{sec-CINT}. The research of T.B. was
partly supported by ANR- 11-LABX-0040-CIMI within the program
ANR-11-IDEX-0002-02 and by ANR contract ANR-14-CE25-0003-0. The research of V.J.
was partly supported by NSERC. The work of C.-A.P. was partly funded by
Excellence Initiative of Aix-Marseille University - A*MIDEX, a French
``Investissements d'Avenir'' programme.
\section{Probabilistic setting}
\label{sec-prob}
\subsection{Variance-gamma distribution of Gaussian vectors}
Let $X$ and $Y$ be independent $\rr^n$-valued Gaussian vectors  with zero mean 
and covariance $M>0$, and let $L>0$ be a matrix on $\rr^n$. The probability 
distribution of the random variable 
\[
Q= X\cdot LX-Y\cdot LY
\]
is called {\sl variance-gamma} distribution of the pair $(L,M)$. We shall denote 
it by $\QQ_\vg$. Obviously,
$$
\QQ_\vg(]-\infty, s])
=\int_{x\cdot Ny\leq s}\d\cN(x)\,\d\cN(y)
$$
where $\cN$ denotes the standard (zero mean, unit covariance) Gaussian measure 
on $\rr^n$ and the matrix $N$ is given by
\beq
N=2L^{1/2}ML^{1/2}.
\label{Ndef}
\eeq

In this subsection we shall compute the density of $\QQ_\vg$  in terms 
of the modified Bessel functions of the second kind
\beq
K_\nu(z)=\int_0^\infty\e^{-z\cosh t}\cosh(\nu t)\d t,
\qquad\left(\nu\in\cc,|\arg z|<\frac\pi2\right),
\label{Knu}
\eeq
and the function
\[
\rr^n\ni x\mapsto\kappa(x)=\left(\sum_{j=1}^n \lambda_j^2 x_j^2\right)^{1/2},
\]
where $0<\lambda_1\leq\lambda_2\cdots \leq \lambda_n$ denote the repeated 
eigenvalues of the matrix~\eqref{Ndef}.

\bep\label{rep-1} The density of $\QQ_\vg$ is given by
\beq 
f_\vg(s)= |s|^{(n-1)/2}\int_{S^{n-1}}K_{(n-1)/2}\left(\frac{|s|}{\kappa(\hat k)}\right)
\frac{\d\sigma(\hat k)}{(2\pi\kappa(\hat k))^{(n+1)/2}},
\label{const}
\eeq
where $\sigma$ denotes the Lebesgue measure on the unit sphere $S^{n-1}$ in 
$\rr^n$.
\eep
{\bf Remark 1.} Although this result is well-known, due to the  lack of a convenient 
reference we give a proof below. 

{\bf Remark 2.} It follows immediately from the definition~\eqref{Knu} that
$K_\nu(x)>0$ for $x>0$ and $\nu\in\rr$. Hence~\eqref{const} implies that
$f_\vg(s)>0$ for $s\not=0$.
For in depth discussion of Bessel functions we refer the reader 
to Watson's treatise~\cite{W}. In this note we shall make use of the following 
facts:
\ben
\item The only singularity of $K_\nu$ is a logarithmic branching point at $z=0$, 
with the leading asymptotics
$$
K_\nu(z)\sim\frac{\Gamma(\nu)}{2}\left(\frac z2\right)^{-\nu},\qquad
\left(\Re\nu>0,z\to0\right).
$$
Moreover, the function $[0,\infty[\ni x\mapsto x^\nu K_\nu(x)$ is monotone
decreasing. It follows that $f_\vg$ is a monotone decreasing function of $|s|$
satisfying
$$
0<f_\vg(s)\le
f_\vg(0)=\frac{\Gamma\left(\frac{n-1}{2}\right)}{4(2\pi)^{(n+1)/2}}
\int_{S^{n-1}}\frac{\d\sigma(\hat k)}{\kappa(\hat{k})}.
$$
\item For fixed $\nu$ the asymptotic expansion
\beq
K_\nu(z)\sim \left(\frac{\pi}{2z}\right)^{1/2}\e^{-z}\sum_{k=0}^\infty \frac{a_k(\nu)}{z^k},\qquad
\left(|z|\rightarrow\infty,|\arg z|<3\pi/2\right)
\label{bessel3}
\eeq
holds with $a_0(\nu)=1$ and 
\[a_{k}(\nu)=\frac{(4\nu^2-1^2)(4\nu^2-3^2)\cdots (4\nu^2-(2k-1)^{2})}{k!\, 8^k}.
\]
\item For half-integer $\nu=m+\frac{1}{2}$, $m\in\nn$, one has
\beq
K_\nu(z)=\left(\frac{2z}{\pi}\right)^{1/2}\,k_m(z).
\label{km}
\eeq
where $k_m$ denotes the modified spherical Bessel function of the second kind
$$
k_m(z)=\frac{\pi}2(-z)^m\left(\frac1z\frac{\d\ }{\d z}\right)^m\frac{\e^{-z}}z.
$$
\een

{\bf Remark 3.} If  $N=\lambda I_n$, where $I_n$ denotes the identity matrix on 
$\rr^n$ and $\lambda>0$, then the integrand in~\eqref{const} is a constant and 
thus
\beq
f_\vg(s)=|S^{n-1}|\,K_{(n-1)/2}\left(\frac{|s|}{\lambda}\right)
\frac{|s|^{(n-1)/2}}{(2\pi\lambda)^{(n+1)/2}},
\label{form}
\eeq
where $|S^{n-1}|=2\pi^{n/2}/\Gamma(n/2)$ is the area of $S^{n-1}$.

{\bf Remark 4.} In the special case $n=2$ and $N=\lambda I_n$ it follows 
from~\eqref{form} and~\eqref{km} that
\beq
f_\vg(s)=\frac{\e^{-|s|/\lambda}}{2\lambda}.
\label{air0}
\eeq
More generally, for $n=2$, \eqref{const} reduces to
$$
f_\vg(s)=
\int_0^{2\pi}\frac{\e^{-|s|/\kappa(\varphi)}}{4\pi\kappa(\varphi)}\d\varphi,
$$
with
$\kappa(\varphi)=\sqrt{\lambda_1^2\sin^2(\varphi)+\lambda_2^2\cos^2(\varphi)}$.
Elementary manipulations lead to

\beq
f_\vg(s)= \frac{\theta}{2\pi}\int_0^\pi 
\sqrt{\frac{1-\varepsilon^2}{1 +\varepsilon \cos\varphi}}
\e^{-\theta|s|\sqrt{1+\varepsilon \cos\varphi}}\d \varphi,
\label{air1}
\eeq
where 
\[
\varepsilon=\frac{\lambda_2^2-\lambda_1^2}{\lambda_2^2 +\lambda_1^2}, \qquad \theta =\left(\frac12\left(\lambda_1^{-2}+\lambda_2^{-2}\right)\right)^{1/2}.
\]

{\bf Proof of Proposition~\ref{rep-1}.}  Setting $U=N^{-1/2}L^{1/2}(X-Y)$ and 
$V=N^{-1/2}L^{1/2}(X+Y)$ one has $Q=U\cdot NV$. Note that  $U$ and $V$ are independent
identically distributed Gaussian random vectors with 
zero mean and unit covariance. It follows that  the characteristic function of 
$\QQ_\vg$ can be written as
\beq
\chi_\vg(\alpha)
=\int_{\rr^{2n}}\e^{\i\alpha x\cdot Ny}\d\cN(x)\,\d\cN(y).
\label{chi}
\eeq
Performing the Gaussian integral over $y$ and invoking the invariance of 
$\cN$ under orthogonal transformations
further yields
$$
\chi_\vg(\alpha)
=\int_{\rr^n}\e^{-\frac{\alpha^2}{2}x \cdot N^2x}\d\cN(x)
=\int_{\rr^n}\e^{-\alpha^2\kappa(x)^2/2}\d\cN(x).
$$
It follows that 
\[
f_\vg(s)
=\int_\rr \e^{-\i\alpha s}\left(
\int_{\rr^n}\e^{-\alpha^2\kappa(x)^2/2}\d\cN(x)\right)
\frac{\d\alpha}{2\pi}.
\]
Performing the Gaussian integral over $\alpha$, we derive
\[
f_\vg(s)= (2\pi)^{-1/2}\int_{\rr^n}
\e^{-s^2/2\kappa (x)^2}\frac{\d\cN(x)}{\kappa(x)},
\]
and passing to the spherical coordinates $r=\|x\|$ and $\hat k=x/\|x\|$ we obtain
\[
f_\vg(s)= (2\pi)^{-(n+1)/2}\int_{S^{n-1}}\left(\int_0^\infty 
\e^{-\frac{1}{2}\left(r^2+r^{-2}s^2\kappa(\hat k)^{-2}\right)}r^{n-2}
\d r\right)
\frac{\d \sigma (\hat k)}{{\kappa(\hat k)}}.
\]
The change of variable $r^2=|s|\e^t/\kappa(\hat k)$ gives
$$
\int_0^\infty 
\e^{-\frac{1}{2}\left(r^2+r^{-2}s^2\kappa(\hat k)^{-2}\right)}r^{n-2}\d r
=\left(\frac{|s|}{\kappa(\hat k)}\right)^{(n-1)/2}
\int_0^\infty\e^{-\frac{|s|}{\kappa(\hat k)}\cosh(t)}\cosh((n-1)t/2)\d t,
$$
and comparison with~\eqref{Knu} yields the result.\hfill\qed

\subsection{Variance-gamma distribution of a Gaussian process}
\label{sec-gau}
Consider a stationary $\rr^n$-valued Gaussian stochastic process 
$X_t=(X_t^{(1)}, \ldots, X_t^{(n)})$, $t\geq 0$,  with mean zero and  covariance 
$M>0$. Given a  matrix $L>0$ on $\rr^n$ we set
\[
Q_t= X_{t}\cdot LX_{t} - X_{0}\cdot L X_{0},
\]
and  denote by $\QQ_t$ the law of $Q_t$. 

Setting again $N=2L^{1/2}ML^{1/2}$ and $U_t=N^{-1/2}L^{1/2}(X_{t}-X_{0})$, 
$V_t=N^{-1/2}L^{1/2}(X_{t}+X_{0})$, one has $Q_t=U_t\cdot NV_t$. The Gaussian 
random vectors $U_t$ and $V_t$ are identically distributed, with zero mean, but they do not need to be independent. To deal with this point 
we introduce the $n\times n$-matrix $\Delta(t)$ with entries
$$
\Delta_{ij}(t)=\langle X_{t}^{(i)}X_{0}^{(j)}\rangle
$$
and formulate
\begin{quote}{\bf Assumption (A)} 
\[
\lim_{t\rightarrow \infty}\Delta(t)=0.
\]
\end{quote}

\bep\label{impo}
Suppose that {\bf (A)} holds. Then  $\QQ_t\to\QQ_\vg$ weakly as $t\to\infty$. 
Moreover, this convergence is accompanied with the following Large Deviations Principle: 
for any open set $O\subset\rr$, 
\beq
\lim_{t\rightarrow \infty}\frac{1}{t}\log \QQ_t(tO)=-\inf_{\theta \in O} I(\theta),
\label{LDP}
\eeq
where the rate function is given by
 \[
 I(\theta)=\frac{|\theta|}{\lambda_n}.
 \] 
\eep
\proof In terms of the $\rr^{2n}$-valued random variable $Z_t=(U_t,V_t)$ one has
$$
Q_t=Z_t\cdot\widetilde{N}Z_t,\qquad
\widetilde{N}=\frac12\left[\begin{matrix}0&N\\N&0\end{matrix}\right].
$$
$Z_t$ is Gaussian with mean zero  and covariance
$$
\widetilde{M}_t=\left[\begin{matrix}I_n-A(t)&B(t)\\
B^T(t)&I_n+A(t)\end{matrix}\right],
$$
where $A(t)=A^T(t)$ and $B(t)=-B^T(t)$ are the $n\times n$ matrices given by
$$
A(t)=N^{-1/2}L^{1/2}(\Delta(t)+\Delta^T(t))L^{1/2}N^{-1/2},\qquad
B(t)=N^{-1/2}L^{1/2}(\Delta(t)-\Delta^T(t))L^{1/2}N^{-1/2}.
$$
The characteristic function of $\QQ_t$ is
$$
\chi_t(\alpha)=\int_{\rr^{2n}}\e^{\i\alpha z\cdot\widetilde{N}z}
\e^{-\frac12z\cdot\widetilde{M_t}z}\frac{\d z}{\det(2\pi\widetilde{M_t})^{1/2}}.
$$
Assumption {\bf (A)} gives that  
\beq \lim_{t\rightarrow \infty}\widetilde{M}_t=I_{2n},
\label{bru}
\eeq
and comparing with~\eqref{chi} we deduce that for any $\alpha\in\rr$,
$$
\lim_{t\rightarrow\infty} \chi_t(\alpha)=\chi_\vg(\alpha).
$$
The weak convergence $\QQ_t\to\QQ_\vg$ then follows from Lévy's continuity theorem.

To prove the LDP, it suffices to consider the case $O=]a,b[$. If $0\in O$, then 
$tO$ is an increasing function of $t$ and since $\QQ_\vg$ has a strictly 
positive density, the weak convergence $\QQ_t\to\QQ_\vg$ implies that
$$
\QQ_{t}(tO)\ge\QQ_{t}(O)\to\QQ_\vg(O)=c>0
$$
for $1\le t\to\infty$. Hence,
$$
0\ge\frac1t\log\QQ_t(tO)\ge\frac1t\log c\to0
$$
and the result follows. So we may assume that $0\not\in O$. We 
consider the case $0<a<b$, the other case is similar.  

From the convergence~\eqref{bru} we infer that for large enough $t>0$ there exists
$0<\delta_t<1/2$ such that
\beq
\lim_{t\to\infty}\delta_t=0,\qquad
(1-\delta_t)I_{2n}\le\widetilde{M}_t\le(1+\delta_t)I_{2n}.
\label{Gend}
\eeq
Denoting by $\cN_{\pm\delta}$ the Gaussian
measure on $\rr^{2n}$ with zero mean and covariance $(1\pm\delta)I_{2n}$
and setting
$$
\QQ_{\pm\delta}(A)=\int_{z\cdot\widetilde{N}z\in A}\d\cN_{\pm\delta}(z),
$$
it follows from~\eqref{Gend} that
$$
3^{-n}\QQ_{-\delta_t}\le\QQ_t\le 3^n\QQ_{+\delta_t}
$$
for large enough $t>0$. Further noticing that
$$
\QQ_{\pm\delta_t}(tO)=\QQ_\vg(\tau_tO),\qquad\tau_t=\frac{t}{1\pm\delta_t},
$$
and $\tau_t/t\to1$ as $t\to\infty$ we conclude that it suffices to show that 
\beq 
\lim_{\tau\rightarrow \infty}\frac{1}{\tau}\log \QQ_\vg(\tau O)
=-\frac{a}{\lambda_n}.
\label{bru-3}
\eeq

We proceed to prove~\eqref{bru-3}.

Obviously, 
\[
\QQ_\vg(\tau O)=\tau\int_a^b f_\vg(\tau s)\d s,
\]
and from the asymptotic expansion~\eqref{bessel3} we infer that there is a
constant $C>0$ such that for $\tau$ large enough, all $s\in [a,b]$, and all 
$\hat k\in S^{n-1}$, 
\beq
C^{-1}\frac{1}{\sqrt\tau}\e^{-\frac{\tau|s|}{\kappa(\hat k)}}
\leq K_{(n-1)/2}\left(\frac{\tau|s|}{\kappa(\hat k)}\right)
\leq C\frac{1}{\sqrt\tau}\e^{-\frac{\tau|s|}{\kappa(\hat k)}}.
\label{never-a}
\eeq
Since $\max_{\hat k\in S^{n-1}} \kappa(\hat k)=\lambda_n$, the  
formula~\eqref{const}  and the upper bound in~\eqref{never-a} give, for large 
$\tau$,
\begin{align*}
\QQ_\vg(\tau O)&\le C\sqrt{\tau}
\left(\int_{S^{n-1}}\frac{\d\sigma(\hat k)}{(2\pi\kappa(\hat k))^{(n+1)/2}}\right)
\int_a^bs^{(n-1)/2}\e^{-\tau s/\lambda_n}\d s\\[6pt]
&\le\left[ C
\left(\int_{S^{n-1}}\frac{\d\sigma(\hat k)}{(2\pi\kappa(\hat k))^{(n+1)/2}}\right)
(b-a)a^{(n-1)/2}\right]\sqrt{\tau}\e^{-\tau a/\lambda_n}
\end{align*}
from which the upper bound
\[
\limsup_{\tau\to \infty}\frac1\tau\log\QQ_\vg(\tau O)
\leq-\frac{a}{\lambda_n},
\]
follows. To prove the corresponding lower bound, fix $0<\epsilon<b-a$ and pick
a neighborhood $\cO_\epsilon$ of $(0,0,\ldots,1)$ in $S^{n-1}$ such that
$$
\inf_{\hat k\in\cO_\epsilon}\kappa(\hat k)\geq\lambda_n-\epsilon.
$$ 
Replacing  integration over $S^{n-1}$ with integration over $\cO_\epsilon$
in~\eqref{const}, the lower bound in~\eqref{never-a} gives
\begin{align*}
\QQ_\vg(\tau O)&\ge C^{-1}\sqrt{\tau}
\left(\int_{\cO_\epsilon}
\frac{\d\sigma(\hat k)}{(2\pi\kappa(\hat k))^{(n+1)/2}}\right)
\int_a^bs^{(n-1)/2}\e^{-\tau s/(\lambda_n-\epsilon)}\d s\\[6pt]
&\ge C^{-1}\sqrt{\tau}
\left(\int_{\cO_\epsilon}
\frac{\d\sigma(\hat k)}{(2\pi\kappa(\hat k))^{(n+1)/2}}\right)
\int_a^{a+\epsilon}s^{(n-1)/2}
\e^{-\tau s/(\lambda_n-\epsilon)}\d s\\[6pt]
&\ge\left[\epsilon C^{-1}
\left(\int_{\cO_\epsilon}
\frac{\d\sigma(\hat k)}{(2\pi\kappa(\hat k))^{(n+1)/2}}\right)
(a+\epsilon)^{(n-1)/2}\right]
\sqrt{\tau}
\e^{-\tau(a+\epsilon)/(\lambda_n-\epsilon)},
\end{align*}
from which we infer
\[
\liminf_{\tau\to\infty}\frac{1}{\tau}\log\QQ_\vg(\tau O)
\geq-\frac{a+\epsilon}{\lambda_n-\epsilon}.
\]
Taking $\epsilon \downarrow 0$ we derive the result.\hfill\qed

\section{Hamiltonian harmonic networks}
\label{sec-det}
\subsection{Setup}
\label{setup}
A Hamiltonian harmonic  network is specified by a quadruple
$\left(G,m,\omega,\cC\right)$, where $G$ is a countable set describing vertices
of the network, $m=(m_x)_{x\in G}$ and $\omega=(\omega_x)_{x\in G}$ are elements
of $\ell^\infty_\rr(G)$ describing the masses and frequencies of the oscillators,
and $\cC$ is a non-negative bounded operator on $\ell^2_\rr(G)$ describing the
coupling between the oscillators. We denote by
$\cC_{xy}=\langle\delta_x,\cC\delta_y\rangle$ the matrix elements of $\cC$
w.r.t.\;the standard basis $\{\delta_x\}_{x\in G}$ of $\ell^2_\rr(G)$.

We identify $m$ and $\omega$ with the induced multiplication operators on
$\ell^2_\rr(G)$ and assume that, as such, $m$ and $\omega$ are strictly 
positive\footnote{The operator $A$ acting on a Hilbert space is 
strictly positive whenever $\langle\phi,A\phi\rangle\ge\lambda\|\phi\|^2$
for all $\phi$ and some $\lambda>0$.}.
The finite energy configurations of the network are described by vectors in the 
Hilbert space 
\[
\cK=\ell_\rr^2(G)\oplus\ell_\rr^2(G)\simeq\ell_\rr^2(G)\otimes\rr^2,
\]
where the pair $(p,q)\in\cK$ describes the momenta and positions of the 
oscillators. The Hamiltonian of the network is
$$
H(p,q)=\sum_{x\in G}\left(\frac{p_x^2}{2m_x}
+\frac{m_x\omega_x^2 q_x^2}2\right) 
+\frac{1}{2}\sum_{x,y\in G}\cC_{xy}q_xq_y,
$$
which is the quadratic form associated to the linear map $H:\cK\to\cK$ defined by 
$$
H=\frac12\left[\begin{matrix}m^{-1}&0\\0&m\omega^2+\cC\end{matrix}\right].
$$
$H$ is a bounded strictly positive operator on $\cK$ so that
$$
\langle \phi, \psi\rangle_H=\langle \phi, H\psi\rangle.
$$
defines an inner product on $\cK$. We denote by $\cK_H$ the Euclidean space
$\cK$ equipped with this inner product.

The Hamilton equation of motion writes $\dot{\phi}=\cL\phi$ with
\[
\cL=\left[\begin{matrix}0&-(m\omega^2+\cC)\\m^{-1}&0\end{matrix}\right].
\]
Note that $\cL$ is a bounded skew-symmetric operator on $\cK_H$. It follows that the finite energy
trajectories of the network are described by the action of the one parameter
group of orthogonal transformations $\rr\ni t\mapsto\e^{t\cL}$ of $\cK_H$.

In the sequel we shall make use of the following  dynamical assumption:

\begin{quote} {\bf Assumption (B)} The spectrum of the operator $\cL$, acting on 
$\cK_H$, is purely absolutely continuous
\end{quote}

It follows from Assumption {\bf (B)} that for all $\phi,\psi\in\cK_H$, 
\beq
\lim_{t\to\infty}\langle\phi,\e^{t\cL}\psi\rangle_H=0.
\label{yes}
\eeq

\subsection{Internal and external work, heat}

We fix a finite subset $G_0\subset G$ and define the total/kinetic/potential 
energy of the corresponding subnetwork by
\begin{align*}
H_{G_0}(p,q)&=\sum_{x\in G_0}\left(\frac{p_x^2}{2m_x}+
\frac{m_x\omega_x^2 q_x^2}2\right)
+\frac12\sum_{x,y\in{G_0}}\cC_{xy}q_xq_y,\\
K_{G_0}(p)&=\sum_{x\in G_0}\frac{p_x^2}{2m_x},\\
V_{G_0}(q)&=\sum_{x\in G_0}\frac{m_x\omega_x^2 q_x^2}2
+\frac12\sum_{x,y\in{G_0}}\cC_{xy}q_xq_y.
\end{align*}
To avoid trivialities, we shall assume that there exists $x\in G_0$ and
$y\in G_0^\mathrm{c}$ such that $\cC_{xy}\not=0$.

The force acting on a vertex $x\in G_0$ decomposes into an internal component
$$
F_x^{\Int}(q)=-m_x\omega_x^2q_x-\sum_{y\in G_0}\cC_{xy}q_y
$$
and an external one
$$
F_x^{\Ext}(q)=-\sum_{y\in G_0^\mathrm{c}}\cC_{xy}q_y.
$$
The work done by the external forces over the time interval $[0,t]$ is
equal to the change in internal energy of the subnetwork
\beq
W^\Ext(t)=\sum_{x\in G_0}\int_0^tF_x^{\Ext}(q(s))\dot q_x(s)\d s
=H_{G_0}(p(t),q(t))-H_{G_0}(p(0),q(0)).
\label{eq-Kinetic}
\eeq
Adding to it the work done by the internal forces we get the total work done
on the subnetwork which equals the change in its kinetic energy:
\begin{align*}
W^\Ext(t)+W^\Int(t)&=\sum_{x\in G_0}\int_0^t\left(F_x^{\Int}(q(s))
+F_x^{\Ext}(q(s))\right)\dot q_x(s)\d s\\
&=K_{G_0}(p(t))-K_{G_0}(p(0)).
\end{align*}
The change in potential energy of the subnetwork is $V_{G_0}(q(t))-V_{G_0}(q(0))$ and it is the opposite of the work done by the internal forces,
$$
-W^\Int(t)=-\sum_{x\in G_0}\int_0^tF_x^{\Int}(q(s))\dot q_x(s)\d s
=V_{G_0}(q(t))-V_{G_0}(q(0)).
$$
Thinking of the subnetwork as an open system connected to a reservoir, we can
interpret Eq.~\eqref{eq-Kinetic} as a statement of the first law of 
thermodynamics: the increase of the internal energy of the subnetwork
equals the amount of heat $\Delta Q_t$ extracted from the reservoir. 

\subsection{The equilibrium case}
\label{sec-ec}
In this subsection we assume that the network is in thermal equilibrium at 
temperature $T>0$. This means that $X_t=(p(t),q(t))$ is a stationary Gaussian 
process over the Hilbert space $\cK$ with mean zero and covariance 
\beq
M=\kB T \left[
\begin{matrix}m&0\\
0&(m\omega^2+\cC)^{-1}
\end{matrix}
\right],
\label{Mequil}
\eeq
where $\kB$ is the Boltzmann constant.

Since 
\[
\langle p_x(t)p_y\rangle
=\langle \e^{t{\cal L}}(\delta_{x}\oplus 0),M(\delta_{y}\oplus 0)\rangle,
\]
Assumption {\bf (B)} implies (recall~\eqref{yes}) that 
$$
\lim_{t\to\infty}\langle p_x(t)p_y\rangle=0.
$$
In the same way one shows that
$$
\lim_{t\to\infty}\langle p_x(t)q_y\rangle
=\lim_{t\to\infty}\langle q_x(t)p_y\rangle
=\lim_{t\to\infty}\langle q_x(t)q_y\rangle=0,
$$
and so  the process $(X_t)_{t\ge0}$ satisfies Assumption~{\bf (A)}.
Applying Proposition~\ref{impo} to the quadratic form 
$K_{G_0}$ we infer that the increase in kinetic energy of the subnetwork
converges in law, as $t\to\infty$, to the variance-gamma distribution
of the pair $(M,K_{G_0})$. One easily shows that
$$
N=2K_{G_0}^{1/2}MK_{G_0}^{1/2}=\kB T\,I_{|G_0|},
$$
and so,  by Remark~3, the limiting law has density~\eqref{form} with 
$\lambda=\kB T$ and $n=|G_0|$. The LDP that accompanies this convergence holds 
with the rate function
\[
I(\theta)=\frac{|\theta|}{\kB T}.
\]

Applying Proposition~\ref{impo} to $H_{G_0}$ we deduce in the same way that
the heat $\Delta Q_t$ absorbed by the subnetwork over the time 
interval $[0,t]$ converges in law, as $t\to\infty$, to the variance-gamma 
distribution of the pair $(M,H_{G_0})$. In this case
$$
N=2H_{G_0}^{1/2}MH_{G_0}^{1/2}=\kB T\,
\left[\begin{matrix}I_{|G_0|}&0\\
0&V_{G_0}^{1/2}V^{-1}V_{G_0}^{1/2}
\end{matrix}
\right],
$$
where $V=m\omega^2+\cC$ and $V_{G_0}$ denote its restriction to $G_0$ with 
Dirichlet boundary conditions on the complement $G_0^{\rm c}$. The 
density of the limiting law is given by~\eqref{const} and, in this case, 
depends on the details of the interaction. It is interesting to notice that,
by Schur's complement formula,
$$
V_{G_0}^{1/2}V^{-1}V_{G_0}^{1/2}
=(I_{|G_0|}-V_{G_0}^{-1/2}\cC V_{G_0^{\rm c}}^{-1}\cC V_{G_0}^{-1/2})^{-1}
>I_{|G_0|}
$$
so that the LDP for heat holds with rate function
$$
I(\theta)=\frac{|\theta|}{(1-\vartheta)\kB T},
$$
where $\vartheta=\|V_{G_0}^{-1/2}\cC V_{G_0^{\rm c}}^{-1}\cC V_{G_0}^{-1/2}\|>0$.

\subsection{The non-equilibrium case}
\label{sec-nec}

Let
\[
G=\bigcup_{a}G_a
\]
be a finite partition of $G$ and
\[
\cK=\bigoplus_a\cK_a
\]
the induced decomposition of Hilbert space $\cK$. We restrict $\cC$ to
$\ell^{2}_\rr(G_a)$ by setting $\langle\delta_x,\cC_a\delta_y\rangle=\cC_{xy}$
if $x,y\in G_a$ and zero otherwise. This leads to (non-interacting) subnetworks
$(G_a,m_a,\omega_a,\cC_a)$. Suppose that the $a$-th subnetwork is in thermal
equilibrium at temperature $T_a$, let $M_a$ be the corresponding covariance and
set
\[
M=\bigoplus_a M_a.
\]
Suppose also that the full network was initially described by a Gaussian random 
field  over $\cK$ with mean zero and covariance $M$. The initial probability 
measure $\PP$ evolves under the interacting dynamics determined by $\cL$ and 
at time $t$, $\PP_t$ is a Gaussian probability measure with mean zero and 
covariance 
\[
M_t=\e^{t\cL}M\e^{t\cL^\ast}.
\]

Set 
\[
\widehat{\cL}=\bigoplus_a\cL_a,
\]
where $\cL_a$ denotes the generator of the dynamics of the $a$-th subnetwork.
Assuming that $\cC-\bigoplus_a\cC_a$ is a trace class operator and that
Condition {\bf (B)} holds, a standard trace-class scattering argument~\cite{LS} 
gives that  the strong limits
\[
W_+=\slim_{t\to\infty}\e^{-t\widehat{\cL}}\e^{t\cL}
\]
exists. Hence,
\[
M_+=\slim_{t\to\infty}M_t
\]
also exists and $M_+=W_+ MW_+^\ast$. The Gaussian measure $\PP_+$ over $\cK$ with 
mean zero and covariance $M_+$ is the non-equilibrium steady state of the network 
associated to  the initial state determined by $M$. Obviously, $\PP_t\to\PP_+$ weakly
as $t\to\infty$.

One can now repeat the analysis of the previous section for the stationary 
Gaussian process $X_t=(p(t), q(t))$ with mean zero and covariance $M_+$.
Both the change in kinetic and total energy of a finite subnetwork converge in
law to a variance-gamma distribution as $t\to\infty$. However, in this case 
there  is no simple  expression for the $\lambda_j$'s. These eigenvalues depend on 
detailed dynamical properties of the model encoded in the scattering matrix 
$W_+$. Even in the simplest example of a harmonic crystal analyzed 
in~\cite[Section 3.2]{JPS2}, not much is known about the $\lambda_j$'s outside 
perturbative regimes.

\section{Stochastic networks}
\label{sec-stoch}
\subsection{Preliminaries}
In this section we shall consider $\rr^n$-valued Gaussian stochastic processes
$(Z_t)_{t\ge0}$ that arise as solutions of SDE's of the form 
\beq
\d Z_t = AZ_t +B\d w_t,
\label{Cauchy}
\eeq
where $A$ and $B$ are  $n\times n$ and $n\times m$ matrices and $w(t)$ a 
standard $\rr^m$-valued Wiener process. The solution of the Cauchy problem~\eqref{Cauchy} with initial condition $Z_0$ is 
$$
Z_t=\e^{tA}Z_0+\int_0^t\e^{(t-s)A}B\d w_s.
$$
We shall assume that $A$ is stable, that is, that all its eigenvalues have
strictly negative real part, and set
\beq
M=\int_0^\infty\e^{sA}BB^\ast\e^{sA^\ast}\d s. 
\label{never}
\eeq
Note that $M$ is non-negative and satisfies the Lyapunov equation 
\[
AM + MA^\ast+BB^\ast=0.
\]
The pair $(A,B)$ is called controllable if the smallest $A$-invariant subspace 
that contains $\Ran B$ is equal to $\rr^n$. If $(A,B)$ is controllable, then 
$M>0$ and the centered Gaussian measure $\P$ on $\rr^n$ with covariance $M$ is 
the unique stationary measure for the process $(Z_t)_{t\ge0}$. Moreover
$$
\Delta_{ij}(t)=\langle Z_t^{(i)}Z_0^{(j)}\rangle=(\e^{tA}M)_{ij},
$$
satisfies Assumption~{\bf (A)}.

We now turn to  two examples  of the above framework: stochastic harmonic
networks and stochastic RC-circuit. 

\subsection{Stochastic harmonic networks}

We start with a Hamiltonian harmonic network 
$\left(G,m_,\omega, \cC\right)$ where the vertex set $G$ is assumed to be 
{\em finite}. To each $x\in G$ we associate two parameters describing 
a thermal reservoir coupled to the oscillator at $x$: the 
temperature $T_x>0$ and the relaxation rate $\gamma_x\geq 0$ ($\gamma_x=0$ 
meaning: no reservoir attached to $x$). The external force acting on the 
oscillator at $x$ has the usual Langevin form 
\[
F_x(p,q)=(2\gamma_x\kB T_x)^{1/2}\dot w_x-\gamma_xp_x,
\]
where the $\dot w_x$ are independent standard white noises. The equation of 
motion of stochastic network is
$$
m_x\dot q_x=p_x,\qquad
\dot p_x=-m_x\omega_x^2q_x-\sum_{y\in G}{\cal C}_{xy}q_y+F_x(p,q).
$$
With $Z_t=(p(t),q(t))$ and obvious notation it can be cast to the 
form~\eqref{Cauchy} with
\[
A=\left[\begin{matrix}-\gamma&-(m\omega^2+\cC)\\
m^{-1}&0\end{matrix}\right], \qquad 
B=\left[\begin{matrix}(2\gamma\kB T)^{1/2}&0\\
0&0\end{matrix}\right].
\]
The matrix $A$ is stable and we shall assume that the pair $(A,B)$ is
controllable. Distributing $Z_0$ according to the centered Gaussian measure with
covariance $M$ given by~\eqref{never} yields a stationary Gaussian process
satisfying  Assumption {\bf (A)}. Note that in  the equilibrium case,
$T_x=T$ for all $x\in G$, $M$ is given by~\eqref{Mequil} which is now a finite
matrix.

The variations of the total/kinetic energy of the network can be analyzed in the
same way as in Subsections~\ref{sec-ec} and~\ref{sec-nec}. In particular, 
in both cases one has the convergence in law to an appropriate variance-gamma 
distribution.  The advantage of the  stochastic setting over the Hamiltonian 
setting described in Section~\ref{sec-det} is that the covariances are 
finite-dimensional matrices and thus accessible to numerical computations. In 
some important examples they can be explicitly computed. In the next subsection 
we analyze one such example. 

\subsection{Stochastic RC-circuit}
\label{sec-CINT}

\begin{figure}
\begin{center}
\includegraphics[width=.5\linewidth]{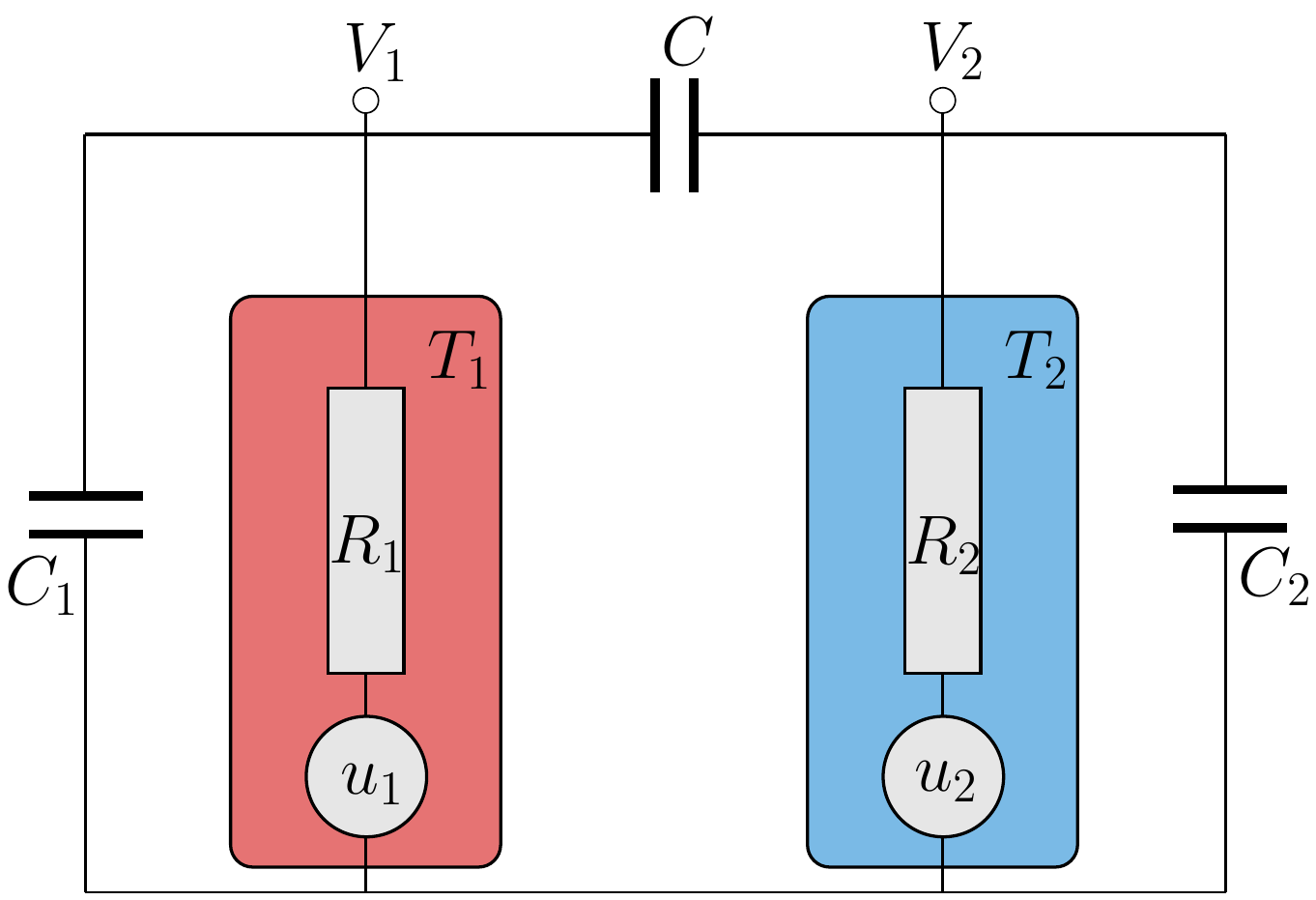}
\end{center}
\caption{\label{circuit} The RC-circuit of~\cite{CINT}: The resistances $R_1$ and 
$R_2$ are kept at temperatures $T_1$ and $T_2$ and are coupled via capacitance 
$C$. $C_1$ and $C_2$ are the inner circuit capacitances. $u_1$ and $u_2$ denote
the fluctuating Johnson-Nyquist voltage. The voltages $V_1$ and $V_2$ 
are amplified by two low noise amplifiers.}
\end{figure}

The RC-circuit subject of the experimental investigations of~\cite{CINT}
is described in Figure~\ref{circuit}. We set 
\[
V=\left[\begin{matrix}V_1 \\V_2\end{matrix}\right],\qquad  
\cC=\left[\begin{matrix}C+C_2 &-C\\-C&C+C_1\end{matrix}\right],\qquad 
\cR=\left[\begin{matrix}R_1 &0\\0&R_2\end{matrix}\right],\qquad 
\cT=2\kB\left[\begin{matrix}T_1 &0\\0&T_2\end{matrix}\right].
\]
The circuit analysis gives that the voltages satisfy the SDE 
\[
\d V_t=AV_t\d t+B\d w_t,
\]
where $A=-\cC^{-1}\cR^{-1}$ and $B=\cC^{-1}(\cT\cR^{-1})^{1/2}$.
One easily verifies that the matrix $A$ is stable and that the pair  $(A,B)$ is controllable.
The stochastic term is due to Nyquist's thermal noise in the resistors.  We denote by $Q_1(t)$ and $Q_2(t)$ the powers dissipated in 
the resistances $R_1$ and $R_2$ at time $t$. They satisfy, in Stratonovich sense,
\begin{align*}
\d Q_1(t)&= V_1(t)((C_1 +C)\d V_1(t)- C\d V_2(t)),\\[2mm]
\d Q_2(t)&= V_2(t)(-\d V_1(t) + (C_2 +C)\d V_2(t)). 
\end{align*}
We are interested in the total dissipated power defined by 
\[
Q_t = Q_1(t) + Q_2(t),
\]
where we set $Q_0=0$. By direct integration,
\[
Q_t =\frac{1}{2}\left(V_t \cdot {\cal C} V_t -V_0 \cdot {\cal C}V_0\right).
\]
 The eigenvalues of 
\[ N= \int_0^\infty {\cal C}^{1/2}\e^{sA}BB^\ast\e^{sA^\ast}{\cal C}^{1/2}\d s
\]
can be computed by diagonalizing $A$ and evaluating the integral. The eigenvalues are given by 
\[
\lambda_{\pm}=\frac\kB2\left(T_1+T_2\pm|T_1-T_2| \sqrt{1-\Lambda^2}\right),
\]
where 
\[
\Lambda= \frac{\sqrt{R_1R_2}\,C}{\frac12(R_1+R_2)C+\frac12(R_1C_1+R_2C_2)}\in]0,1[.
\]

Recalling Remark 4 after Proposition~\ref{rep-1}, we can now complete the 
theoretical analysis of the total dissipated power.
\begin{figure}[h!]
\begin{center}
\includegraphics[width=.48\linewidth]{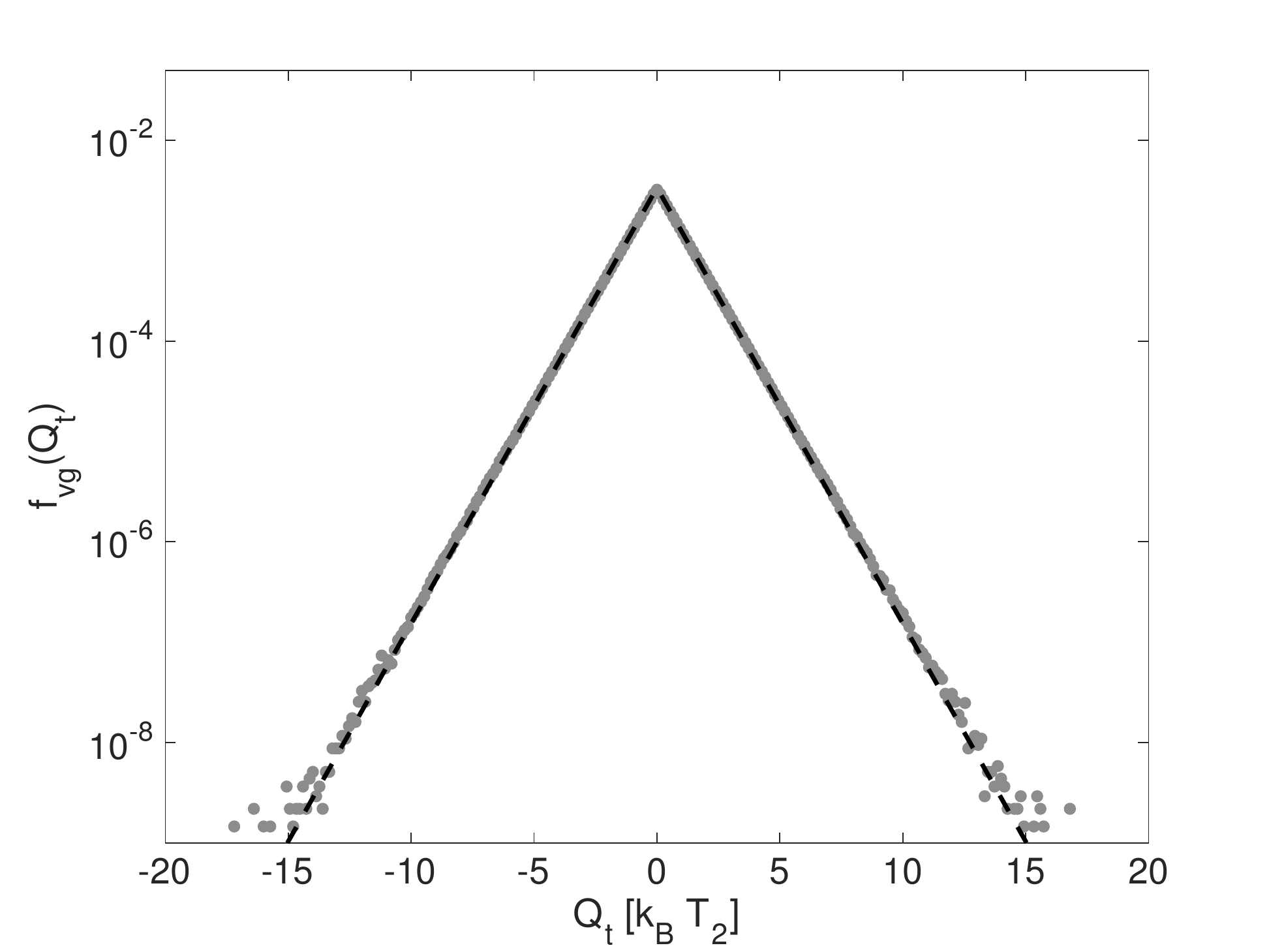}
\includegraphics[width=.48\linewidth]{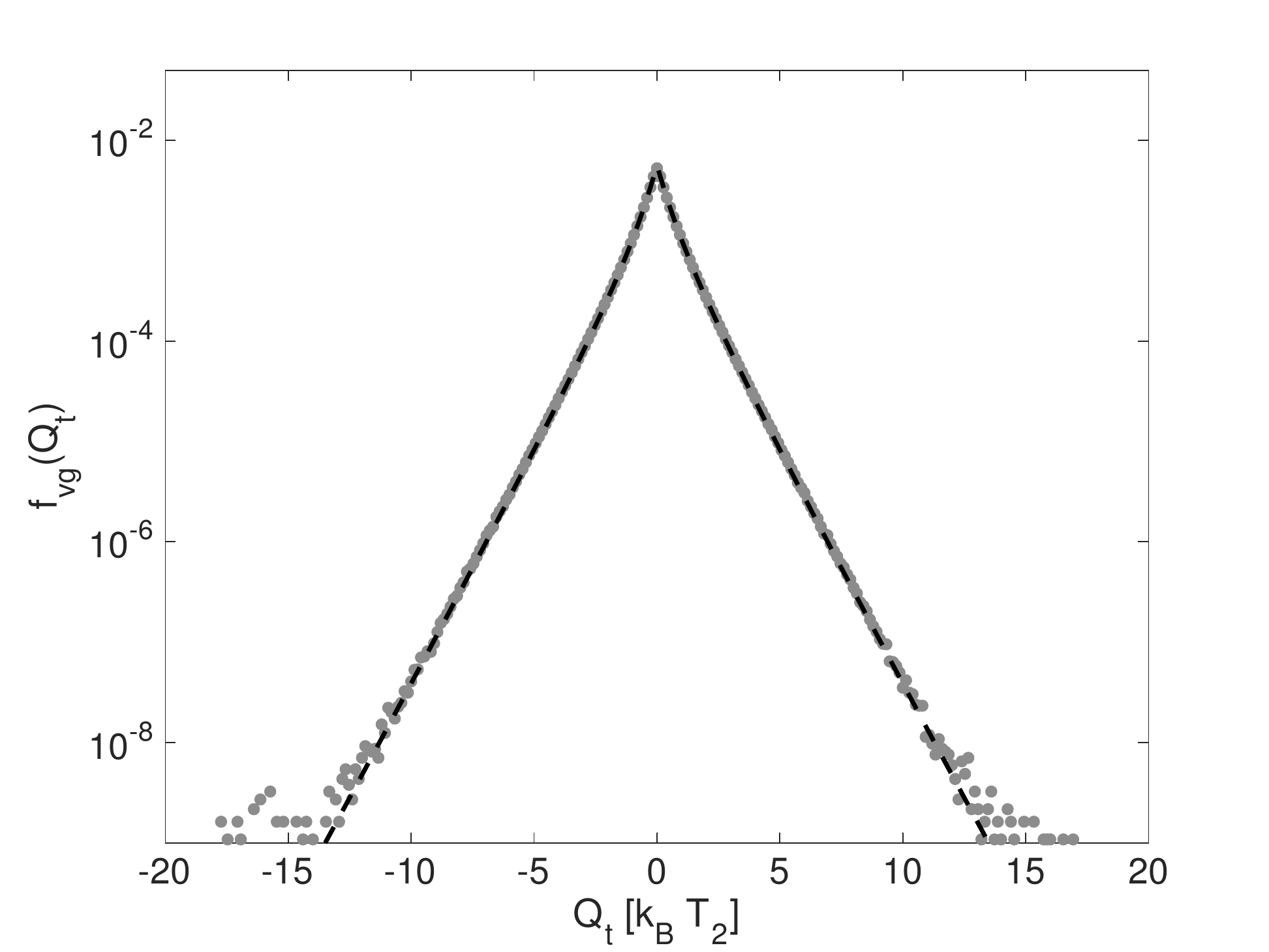}
\end{center}
\caption{\label{fig:CINT_graph} Empirical and theoretical distribution of $Q_t$
with the experimental parameters of~\cite{CINT}: $R_1=R_2=10^8\ \Omega$,
$C=10^{-10}\ F$, $C_1=6.8\times 10^{-10}\ F$ and $C_2=4.2\times 10^{-10}\ F$.
The heat units are in multiples of $\kB T_2$. For the empirical distribution
$t=0.2$ s, the theoretical distribution  is the limiting $t\rightarrow\infty$
variance-gamma distribution\protect\footnotemark. Both at equilibrium (left
pane, $T_1=T_2=296K$) and out of equilibrium (right pane, $T_1=88K$,
$T_2=296K$), the theoretical prediction (black dashed lines) matches almost
perfectly the empirical distribution computed from the experimental data (grey
dots). Note that we have taken the theoretical prediction as a reference for the
normalization of the empirical distribution. The discrepancy between the
theoretical and empirical distribution for $|Q_t|\geq 10\kB T_2$ is due to the
lack of experimental data in this energy range.}
\end{figure}
\footnotetext{Since the largest eigenvalue of $A$ is smaller than $-12$ and $\exp(-.2\times 12)\leq .091$, the experimental data is comparable to the limit $t\to\infty$.}
\begin{quote}
{\bf Thermal equilibrium case} $T_1=T_2=T$. Then $\lambda_-=\lambda_+=\kB T$ and
it follows from Proposition~\ref{impo} and Eq.~\eqref{air0} that $Q_t$ converges
in law  as $t\to\infty$ to the probability distribution with density
\[
f_{\rm eq}(s)=\frac{\e^{-|s|/\kB T}}{2\kB T}.
\]
\end{quote}

\begin{quote}
{\bf Non-equilibrum case}  $T_1<T_2$. Then $\kB
T_1<\lambda_-<\frac\kB2(T_1+T_2)<\lambda_+<\kB T_2$ and Proposition~\ref{impo}
and Eq.~\eqref{air1} yield that $Q_t$ converges in law as $t\to\infty$ to the
probability distribution with density
$$
f_{\rm neq}(s)=\frac{\theta}{2\pi}\int_0^\pi
\sqrt{\frac{1-\varepsilon^2}{1 +\varepsilon \cos\varphi}}
\e^{-\theta|s|\sqrt{1+\varepsilon \cos\varphi}}\d\varphi,
$$
where 
\[
\varepsilon
=\frac{\lambda_+^2-\lambda_-^2}{\lambda_+^2 +\lambda_-^2}
=\frac{\sqrt{1-\Lambda^2}\,|T_1^2-T_2^2|}{(T_1^2 + T_2^2)-\frac12\Lambda^2(T_1-T_2)^2},
\]
\[
\theta=\left(\frac12\left(\lambda_-^{-2}+\lambda_+^{-2}\right)\right)^{1/2}
=\frac1\kB\frac{\sqrt{\frac12(T_1^2+T_2^2)-\frac{\Lambda^2}{4}(T_1-T_2)^2}}
{T_1T_2+\frac{\Lambda^2}{4}(T_1-T_2)^2}.
\]
\end{quote}
Figure~\ref{fig:CINT_graph} summarizes the comparison of these theoretical predictions with the experimental data from~\cite{CINT}.

\end{document}